# Attention-Enhanced Cross-Task Network for Analysing Multiple Attributes of Lung Nodules in CT

Xiaohang Fu, Lei Bi, Ashnil Kumar, Michael Fulham, and Jinman Kim

X. Fu, L. Bi, M. Fulham, and J. Kim are with the School of Computer Science, Faculty of Engineering, The University of Sydney, NSW 2006, Australia (e-mail: xiaohang.fu@sydney.edu.au, lei.bi@sydney.edu.au, michael.fulham@sydney.edu.au, and jinman.kim@sydney.edu.au). M. Fulham is also with the Department of Molecular Imaging, Royal Prince Alfred Hospital. A. Kumar is with the School of Biomedical Engineering, Faculty of Engineering, The University of Sydney, NSW 2006, Australia (email: ashnil.kumar@sydney.edu.au).

Corresponding authors: Xiaohang Fu and Jinman Kim.

# Abstract

Accurate characterisation of visual attributes such as spiculation, lobulation, and calcification of lung nodules is critical in cancer management. The characterisation of these attributes is often subjective, which may lead to high inter- and intra-observer variability. Furthermore, lung nodules are often heterogeneous in the cross-sectional image slices of a 3D volume. Current state-of-the-art methods that score multiple attributes rely on deep learning-based multi-task learning (MTL) schemes. These methods, however, extract shared visual features across attributes and then examine each attribute without explicitly leveraging their inherent intercorrelations. Furthermore, current methods either treat each slice with equal importance without considering their relevance or heterogeneity, which limits performance. In this study, we address these challenges with a new convolutional neural network (CNN)-based MTL model that incorporates multiple attention-based learning modules to simultaneously score 9 visual attributes of lung nodules in computed tomography (CT) image volumes. Our model processes entire nodule volumes of arbitrary depth and uses a slice attention module to filter out irrelevant slices. We also introduce cross-attribute and attribute specialisation attention modules that learn an optimal amalgamation of meaningful representations to leverage relationships between attributes. We demonstrate that our model outperforms previous state-of-the-art methods at scoring attributes using the well-known public LIDC-IDRI dataset of pulmonary nodules from over 1,000 patients. Our model also performs competitively when repurposed for benign-malignant classification. Our attention modules also provide easy-to-interpret weights that offer insights into the predictions of the model.

# 1. Introduction

Lung cancer is the leading cause of cancer-related mortality worldwide and accounted for over 18% of the 9.5 million cancer deaths in 2018 [1]. The planning and optimisation of treatment depends on accurate lung nodule characterisation and staging. Computed tomography (CT) is an indispensable tool for the clinical assessment and profiling of lung nodules [2]. Lung nodules can vary in appearance and size, and some visual characteristics are suggestive of cancer [3]. CT has adequate resolution to depict characteristics such as sphericity, spiculation, and calcification, which are referred to as *attributes* in this study. They characterise high-level appearance of nodules at the object level that are more abstract, rather than low-level features such as colour and edges. These attributes are used to classify the malignancy (cancerous or benign) and subtype of nodules [3-5]. Furthermore, attributes may be correlated with other data including clinical reports, and can be used in guiding patient management [4].

The characterisation of high-level attributes of lung nodules from CT images is non-trivial as it involves analysing a 3D stack of image slices where the nodule's visual appearance changes across each image (as shown in Fig. 1). These nodules exhibit different slice thicknesses from different image acquisition setups, and different numbers of image slices for each nodule. Visual features can often seem ambiguous and vague, which exacerbates the subjectivity of this task, and leads to inaccuracies and high inter- and intra-observer variability [6, 7]. Furthermore, the accurate assessment of nodules involves simultaneously considering multiple visual features, which is complex and may be influenced by the experience of the radiologist [6]. A robust automated method may assist in overcoming the limitations of manual assessment.

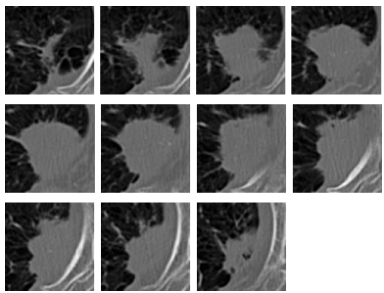

Fig. 1. Sections of a transaxial CT image showing the extent of a lung nodule using soft tissue windows. Individual image slices show the irregular outline of the nodule abutting the lateral pleura.

Earlier methods were mainly directed at scoring or classifying individual attributes, especially malignancy, independently. Various techniques such as a bag-of-frequencies descriptor [8], patch-based handcrafted features (including Scale-Invariant Feature Transform [SIFT] and Histogram of Oriented Gradients [HOG]) [9], and decision trees [10] have been applied. More recent methods are based on deep learning (DL) using convolutional neural networks (CNNs), and multi-task learning (MTL) has been the framework of choice for simultaneously scoring multiple attributes of lung nodules [11-14]. In MTL, relationships between attributes are implicitly leveraged through supervised co-training across multiple tasks. This strategy assumes that the different attributes can be derived from overlapping visual information, so a single model can classify multiple attributes. The earlier stages of the model compute collective features that are shared between the different tasks, while the later stages become more task specific. Chen et al. [11] reported a model that combined features extracted from a CNN and stacked denoising autoencoder, with handcrafted Haar-like and HOG features. They also showed that MTL outperforms single task methods at rating lung nodule attributes. Liu et al. [12] reported Multi-Task deep model with Margin Ranking loss (MTMR-Net), which leveraged the Siamese network architecture with a 152-layer ResNet [15] backbone to simultaneously score attributes and classify nodules as benign or malignant.

The wide variation of lung nodule volume sizes hinders the application of 3D approaches, so existing methods typically adopt a 2D approach which produces classifications on each slice, followed by averaging across the slices of each nodule. However, some slices contain more useful information while others are less useful and may appear misleading. This can impair the classification accuracy of the conventional approach, which analyses slices individually. Additionally, inter-attribute relationships have only been investigated implicitly by previous methods that adopt an MTL framework, typically by sharing CNN features across attributes. This approach, however, does not expose correlations between attributes. Furthermore, studies were limited to the relations between malignancy and the other attributes, while the rest were not explored. Malignancy can be inferred by certain patterns of morphological characteristics such as irregular, lobulated or spiculated margins; while popcorn, diffuse, central, and laminated calcification patterns are usually associated with benign nodules [4, 5, 16, 17]. A previous approach first predicted these characteristics, then used the predictions or intermediate CNN features to infer malignancy [18, 19]. Liu et al. [12] used a model that, with further processing via recursive feature elimination and logistic regression, could score the importance of each attribute towards malignancy.

In DL-based MTL, attention mechanisms that explore cross-task relations have been reported [20]. These approaches learn weights to emphasise elements of intermediate feature vectors by extracting and highlighting more relevant information regarding the present task. Misra et al. [21] reported cross-stitch units to learn optimal linear combinations of shared and task-specific features between different parallel tasks. Other studies have built upon these cross-task connections [22, 23]. Liu et al. [24] used soft attention units that learn task-specific features from a global pool. Zhao et al. [25] used attention-based modules for task-task knowledge transfer, task-feature dependence, and feature-feature interactions to explore relationships between tasks. Coppola et al. [26] presented a gating mechanism to understand the similarities between multiple attributes in skin cancer diagnosis. They formulated the relationship between two attributes as the proportion of shared features. For more abstract clinical attributes, however, there is not a method to expose cross-task interdependencies in a format that encapsulates high-level content and allows for direct understanding of the relationships.

In this study, we propose a CNN-based deep MTL model to simultaneously score multiple attributes in lung nodules. Our model manages variations in size and slice thickness, and extracts and aggregates features from all cross-sections simultaneously for each nodule, yielding a single prediction set. We introduce three attention modules to enhance the accuracy of the model and help elucidate the mechanisms behind predictions. The first module deals with inconsistencies in visual appearances through an image volume by assigning importance to slices containing more useful attribute information and de-emphasising slices in which attributes appear ambiguous or misleading. Our other two attention modules explore the inter-relatedness between attributes explicitly, by defining the relationships as readily interpretable weights. We build on the standard MTL configuration by cross-pollinating between the meaningful and attribute-specific features in the later stages of the model.

Our contributions to the state-of-the-art are:

a) A DL-based MTL model with attention modules that processes entire image volumes of arbitrary depth and scores multiple nodule attributes simultaneously. The importance of each slice is proportionally accounted for via an attention module that decreases the influence of less important slices.

b) A cross-task feature learning attention module to explicitly use inter-attribute dependencies via the combination of high-level CNN representations, and reveal such relationships in an easy-to-understand format that can be used for clinical interpretation.

c) An attribute specialisation attention module to ensure that the high-level CNN representations are meaningful to the attributes.

# 2. Methods

Our proposed model is an MTL framework that scores attributes (subtlety, internal structure, calcification, sphericity, margin, lobulation, spiculation, texture, and malignancy) of a nodule simultaneously. The model has several components, which are illustrated in Fig. 2. The initial stages are concerned with the extraction of deep visual features from the input images that are shared across all attributes. The product of this common pathway is a feature vector that characterises the content of the input stack. Subsequent components further process the nodule content vector in branches that specialise to individual attributes. Given a stack of 2D image slices of the nodule, a single score for each attribute is predicted for the input nodule. All the slices of a 3D nodule are processed simultaneously by arranging slices to be in the batch dimension of the input. In this way, the model supports arbitrary slice numbers and thicknesses without needing interpolation or padding in the axial dimension. Our model regresses floating-point scores for each attribute, as was done by previous studies [11, 12] to accommodate for inter-rater variability in the ground truth ratings.

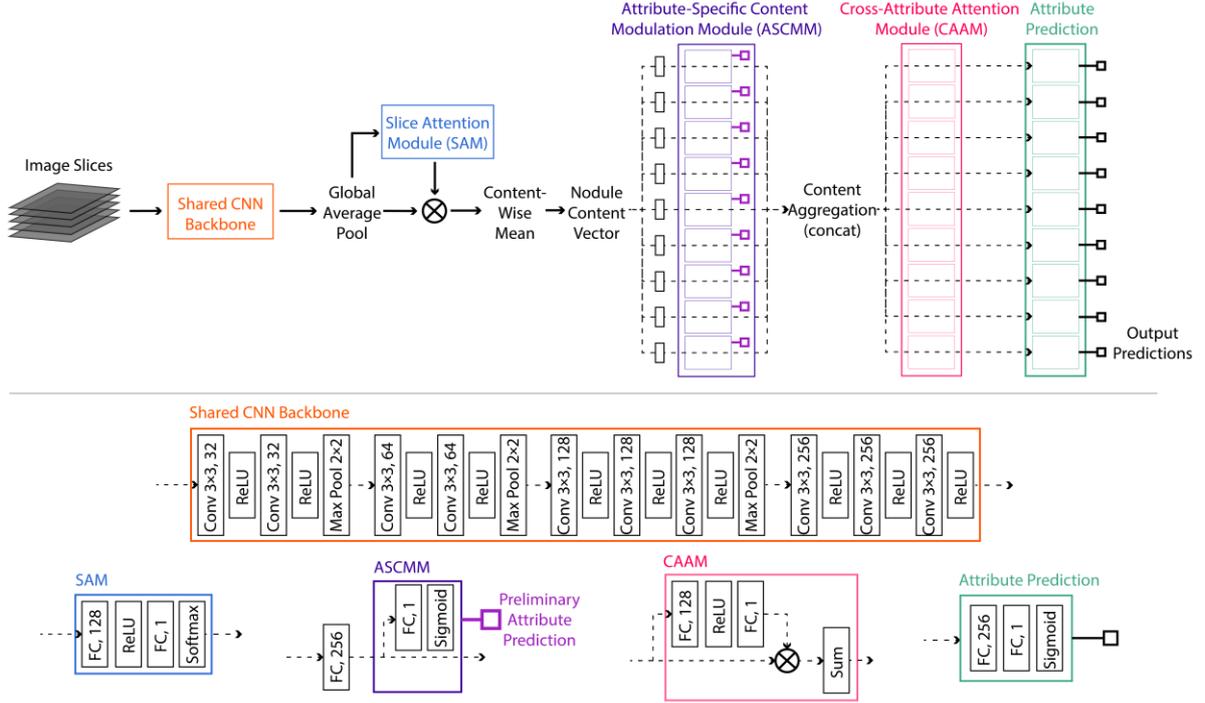

Fig. 2. Our proposed deep learning model with attention modules, and expanded diagrams of the components (best viewed in colour). A high-resolution version is included in the Supplementary Materials.

## 2.1. Shared Feature Extraction

The first component of the model is a CNN backbone (coloured orange in Fig. 2) that extracts deep convolutional features from the input images through several sequential blocks of $3 \times 3$ convolutions, ReLU [27], and max pooling operations. In this component, the image slices of an input volume are regarded as 2D by the CNN. One feature volume with 256 channels per slice is produced, i.e., an image volume of $x \times y \times z$ is input as $z \times x \times y \times 1$, and the backbone outputs $z \times h \times w \times d$ (where $d$ is 256). The CNN automatically learns to discern general visual features that will be leveraged by the scoring tasks downstream.

Immediately after the CNN, we use global average pooling (GAP) [28] to vectorise the feature volumes, reducing the dimensionality from $h \times w \times d$ to $1 \times 1 \times d$, by averaging over each feature channel. GAP is an established technique used in image-based DL models (such as ResNets [15]) for producing more meaningful feature vectors. Elements of the vector may be interpreted as confidence in the different convolutional features. GAP reduces overfitting compared to a simple flatten operation as it yields a shorter vector, thus subsequent fully connected layers will have fewer parameters. GAP is also robust to spatial translations of the input.

## 2.2. Attention Modules

### 2.2.1. Slice Attention

In a nodule volume, some image slices offer greater utility for scoring attributes. As our approach involves considering the entire stack of images of a nodule at one time, it is beneficial to determine which slices contain more useful information and scale their influence on predictions accordingly, rather than simply assume that all slices are equally important.

We propose the slice attention module (SAM) to automatically determine the influence of each slice in a stack of arbitrary $z$-size. The SAM learns to place more weight on slices which contain more relevant information regarding the attributes.

In our model, SAM (coloured blue in Fig. 2) is placed after the GAP layer. The feature vectors for each slice produced by the GAP layer are fed to SAM, yielding a scalar weight per slice according to the following equations:

$$\boldsymbol{q_{SAM}}(\boldsymbol{x_g}) = \boldsymbol{W_1}\Psi(\boldsymbol{W_0}\boldsymbol{x_g} + \boldsymbol{b_0}) + \boldsymbol{b_1} \tag{1}$$

$$\text{let } \boldsymbol{p} = \boldsymbol{q_{SAM}}(\boldsymbol{x_g})$$

$$\alpha(\boldsymbol{p})_i = \frac{e^{p_i}}{\sum_j e^{p_j}} \tag{2}$$

where $\boldsymbol{\alpha} \in \mathbb{R}^{1 \times M}$ denotes the SAM weights for a nodule with $M$ slices, $\boldsymbol{p} \in \mathbb{R}^{1 \times M}$ denotes the pre-softmax SAM weights, and $\boldsymbol{x_g} \in \mathbb{R}^{M \times 256}$ is the SAM input, namely the GAP feature vectors for each slice. SAM is characterised by a set of trainable weights and biases of two fully connected layers, comprising of $W_0 \in \mathbb{R}^{128 \times 256}$, $W_1 \in \mathbb{R}^{128 \times 1}$, $b_0 \in \mathbb{R}^{128 \times 1}$, and $b_1 \in \mathbb{R}^{1 \times 1}$. $\Psi$ denotes the ReLU activation between the two fully connected layers. The softmax function ensures that all SAM weights per stack sum to 1. Thus, the weights indicate the relative importance of each image slice of a nodule.

Subsequently, we use a content-wise weighted sum to aggregate all the slice vectors into a single 256-dimensional vector that characterises the content of each image stack, based on the weights produced by SAM:

$$f_c = \sum_i (\alpha_i x_{g,i}) \tag{3}$$

where $f_c \in \mathbb{R}^{1 \times 256}$ is the content vector of the nodule.

### 2.2.2. Attribute-Specific Content Modulation and Cross-Attribute Attention

As in typical general MTL frameworks, the latter part of our model branches into task-specific pathways and becomes more specialised for individual attributes. However, we hypothesise that it is beneficial to increase flexibility in mixing more specialised features between tasks. To this end, we alter the traditional task specialisation process by introducing two cooperative modules: an attribute-specific content modulation module (ASCMM, purple in Fig. 2) and cross-attribute attention module (CAAM, pink in Fig. 2), to perform cross-attribute feature learning. There is one ASCMM and CAAM unit per attribute.

The ASCMM prompts the specialisation of each attribute pathway via the utilisation of an auxiliary score regression and mean squared error (MSE) loss during training:

$$L^{ASCMM} = \frac{1}{N} \sum_i || \hat{y}_i - y_i^{ASCMM} ||_2^2 \tag{4}$$

where $N$ is the total number of nodules in the training set; $\hat{y}_i \in \mathbb{R}^{1 \times K}$ and $y_i^{ASCMM} \in \mathbb{R}^{1 \times K}$ are the ground truth and auxiliary score predictions of all attributes, with $K = 9$, corresponding to the 9 attributes. This facilitates the production of a content vector that is more specific to each attribute. During inference, the ASCMM is not used, and auxiliary predictions do not contribute towards the final predictions of the model.

After the specialisation of the content vectors for each task, we use CAAM to cross-pollinate content between branches to leverage useful inter-attribute relationships. Each CAAM unit learns the importance of *all* attributes towards the score regression for its own attribute:

$$q_{CAAM}^t(x_s) = W_1^t \Psi(W_0^t x_s + b_0^t) + b_1^t \tag{5}$$

where $q_{CAA}^t \in \mathbb{R}^{1 \times 9}$ denotes the CAAM weights for attribute $t$, and $x_s \in \mathbb{R}^{9 \times 256}$ is the input comprising of the specialised content vectors from all attributes. The CAAM unit for attribute $t$ is characterised by a set of trainable weights and biases of two fully connected layers, comprising of $W_0^t \in \mathbb{R}^{128 \times 256}$, $W_1^t \in \mathbb{R}^{128 \times 1}$, $b_0^t \in \mathbb{R}^{128 \times 1}$, and $b_1^t \in \mathbb{R}^{1 \times 1}$. $\Psi$ denotes the ReLU activation between the two fully connected layers.

Next, the weights produced by the CAAM unit are used via a weighted sum to combine representations from the content vectors of all attributes into a single 256-dimensional vector:

$$f_s^t = \sum_k (q_{CAAM,k}^t x_{s,k}) \tag{6}$$

where $f_s^t \in \mathbb{R}^{1 \times 256}$ is the final output CAAM feature vector for attribute $t$.

## 2.3. Attribute Score Regression and Training

The final component of the model is output prediction (Fig. 2 in green), and this consists of a unique block of two fully connected layers per attribute. Sigmoid activation ensures that output regression scores are normalised within [0, 1].

To train the model we minimise the MSE loss between predicted and ground truth scores, $L^P$:

$$L^P = \frac{1}{N} \sum_i || \hat{y}_i - y_i^P ||_2^2 \tag{7}$$

where $\hat{y}_i \in \mathbb{R}^{1 \times K}$ and $y_i^P \in \mathbb{R}^{1 \times K}$ are the ground truth and final score predictions of all attributes. The final regression loss value that we minimise via standard backpropagation is the sum of the losses from the ASCMM and output components:

$$L^{total} = \lambda L^{ASCMM} + L^P \tag{8}$$

where $\lambda$ is the scalar hyperparameter which balances the ASCMM loss relative to the prediction loss.

# 3. Experiments

## 3.1. Materials

We used the publicly available LIDC-IDRI dataset [7] to evaluate our model. The dataset comprises CT scans of nodules from 1,010 patients. The scans were acquired with a range of scanner models and imaging parameters, obtained from several medical centres in the United States. The nodules were identified and delineated by multiple radiologists. We used nodules that were marked by at least one radiologist with a diameter greater than 3 mm and were further rated for attributes (from 875 patients), as done by comparable studies [11, 12].

Nine high-level attributes were independently scored by up to 4 radiologists per nodule (i.e., not every nodule was scored by 4 radiologists). The attributes were rated from 1 to 5 inclusive, except internal structure (scored 1 to 4) and calcification (scored 1 to 6). The value of attribute rating generally related to the prominence of the attribute in the images (Fig. 3).

For example, a rating of 6 for sphericity indicate that the nodule was very round; a score of 6 for subtlety indicated the nodule was very clearly apparent in the image.

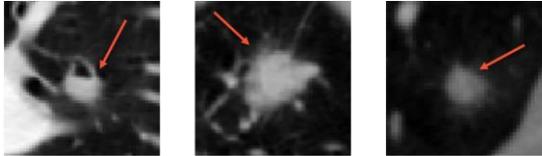

| | | | | | |
|---|---|---|---|---|---|
| **Subtlety** | Fairly Subtle | 3 | Obvious | 5 | Fairly Subtle | 3 |
| **Internal Structure** | Fluid | 2 | Soft Tissue | 1 | Soft Tissue | 1 |
| **Calcification** | Absent | 6 | Non-central | 4 | Absent | 6 |
| **Sphericity** | Ovoid | 3 | Ovoid | 3 | Ovoid/Round | 4 |
| **Margin** | Near Sharp | 4 | Sharp | 5 | Nearly Poorly Defined | 2 |
| **Lobulation** | Nearly No Lobulation | 2 | Nearly No Lobulation | 2 | Medium | 3 |
| **Spiculation** | No Spiculation | 1 | Medium | 3 | Nearly No Spiculation | 2 |
| **Texture** | Solid/Mixed | 4 | Solid | 5 | Non-Solid/Mixed | 2 |
| **Malignancy** | Highly Unlikely | 1 | Moderately Suspicious | 4 | Indeterminate | 3 |

Fig. 3. Sections of CT images showing 3 lung nodules from different patients using soft tissue windows and their attribute ratings by radiologists. The images were cropped to the nodule ROIs and resized to the same spatial resolution. Ratings were scored on a scale of 1 to 5, except for internal structure (1 to 4) and calcification (1 to 6).

There was considerable variability between radiologists' scores and in the number of rating instances per nodule, so we used the average radiologist scores for each attribute as the ground truth, as adopted by comparable studies [11, 12]. Ground truth ratings were normalised to be within [0, 1] for model training.

For image processing, we used the *pylidc* Python package [29] that was created for the LIDC-IDRI dataset. We extracted the volume of interest (VOI) of each nodule based on the common bounding box of the radiologists' annotations that was computed via a consensus consolidation at 50% agreement level. The VOIs were adaptively padded to be square in the lateral *x-y* plane such that the larger diameter in *x* or *y* was 80% of the final lateral dimension. We preserved the anisotropic resolutions between the *x-y* and *z* dimensions as we found that isotropic conversion provided no benefit to prediction error. Only slices that contained nodule-positive pixels as labelled in the annotations were considered; the number of slices ranged from 1 to 45 with a mean of 5.6 (distribution presented in Fig. 4). The annotated bounding boxes had a mean height and width of 15.8 pixels, ranging from 5 to 98 and 5 to 86 pixels, respectively. All images were bilinearly interpolated to $64 \times 64$ in lateral resolution. Image intensities were converted to Hounsfield units. There was a total of 2,622 nodules.

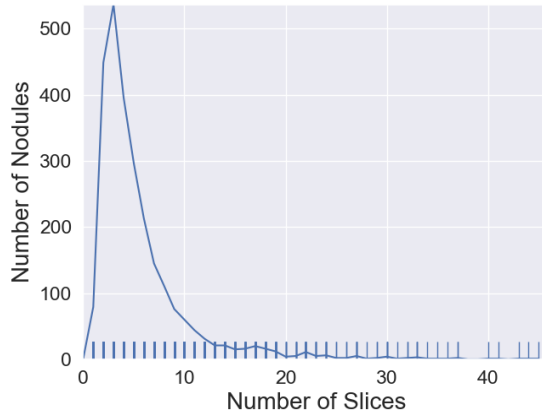

Fig. 4. Number of image slices per nodule in the dataset.

## 3.2. Implementation Details

We kept the following implementation and hyperparameter choices consistent for all experiments to ensure a fair comparison. The networks were trained end-to-end from scratch for 150 epochs using stochastic gradient descent with a nodule batch size of 1 (all the slices for a single nodule at a time). We employed the Adam optimiser [30] to minimise the total multi-task MSE loss at a fixed learning rate of 0.0001, with a first moment estimate of 0.9, second moment estimate of 0.999, and a weight decay constant of 0.0001. We set the weight of the auxiliary loss $\lambda$ to be 0.1, as determined empirically (evaluation in Supplementary Materials Table S1). Convolutional filter weights were initialised using He et al.'s method [31] while biases were initialised to zero. Glorot (Xavier) initialisation [32] was used for fully connected layers. Dropout was not used in any experiment.

Each image was mean-subtracted and normalised to unit variance (using the training set mean and standard deviation). We employed standard online (on-the-fly) image data augmentation by randomly applying a flip (horizontal or vertical), rotation (of 90, 180 or 270 degrees), or reversing the $z$-order of slices, to the input volume. For flips and rotations, identical transformations were used for all slices in a volume. The order of training examples was shuffled every epoch. All networks were implemented using the PyTorch framework [33]. Both training and testing were performed with a 12GB NVIDIA GTX Titan X GPU. Training required about 3 hours for completion.

## 3.3. Evaluation Setup

We performed an ablation study to determine the contributions of each module. We used an attention-free baseline model in which SAM, ASCMM, and CAAM were removed from our model. We evaluated the performance of this baseline, and its performance with the incorporation of only SAM, only CAAM, or both ASCMM and CAAM.

Furthermore, we assessed the performance of the 3D variant of our attention-free baseline. For this model, each input image volume was padded in the axial dimension such that the input size was $64 \times 64 \times 64$ pixels. This was done to accommodate the volume with the largest depth (45 slices), and extends the model to 3D whilst maintaining the same architectural parameters in the extra dimension.

A 5-fold cross-validation was performed for each experimental setup. The nodules were randomly divided into training and testing sets with an 80/20 percent split, resulting in 2098 or 2087 samples for training and 524 or 525 samples for testing. Identical nodule splits were used for each method and we ensured that no nodule was in both the training and test sets of a fold.

Our main performance metric was the mean absolute error (MAE) per attribute between the average radiologist scores and predictions, as adopted by previous studies:

$$MAE = \frac{1}{N} \sum_i |\hat{y}_i - y_i^P| \qquad (9)$$

where $N$ is the total number of nodules in the training set; $\hat{y}_i \in \mathbb{R}^{1 \times K}$ and $y_i^P \in \mathbb{R}^{1 \times K}$ are the ground truth and final score predictions of all attributes, with $K = 9$, corresponding to the 9 attributes. All our computations during testing were based on de-normalised scores, according to original rating scales. The best model per experiment was that which scored the lowest in average MAE across all nodules and attributes. We benchmarked the proposed model to the state-of-the-art methods of Chen et al. [11] and Liu et al. [12], and against methods designed for benign-malignant classification [34-39].

# 4. Results

## 4.1. Ablation Study

The ablation study (Table 1) revealed that prediction performance improved when each attention module was integrated with the attention-free version of the proposed model. Overall, prediction errors with the attention modules were significantly ($p$<0.05) lower than the attention-free baseline, as determined by $t$-tests which compared the absolute errors of the predictions. Performance on 'calcification' improved by the largest margin compared to other attributes.

Table 1. Prediction performance per attribute with or without our attention modules.

| Method | Performance (Mean Absolute Error) | | | | | | | | | |
|---|---|---|---|---|---|---|---|---|---|---|
| | **Sub** | **IS** | **Cal** | **Sph** | **Mar** | **Lob** | **Spi** | **Tex** | **Mal** | **Mean** |
| No Attention | 0.741 | 0.031 | 0.228 | 0.542 | 0.610 | 0.483 | 0.450 | 0.498 | 0.510 | 0.455 |
| SAM | **0.658** | 0.026 | 0.165 | **0.489** | **0.549** | **0.460** | **0.423** | 0.409 | **0.457** | 0.404* |
| CAAM | **0.664** | 0.024 | 0.156 | 0.494 | **0.549** | 0.462 | 0.427 | 0.411 | **0.457** | 0.405* |
| CAAM+ASCMM | 0.671 | **0.022** | **0.146** | 0.490 | 0.550 | **0.454** | **0.423** | **0.408** | **0.453** | **0.402*** |
| SAM+CAAM+ASCMM | **0.664** | **0.020** | **0.141** | **0.484** | **0.536** | **0.460** | **0.424** | **0.399** | **0.453** | **0.398*** |

\* = *p*-value < 0.05 compared to the No Attention baseline
Sub = Subtlety, IS = Internal Structure, Cal = Calcification, Sph = Sphericity, Mar = Margin, Lob = Lobulation, Spi = Spiculation, Tex = Texture, Mal = Malignancy
Best results are indicated by red, second-best by blue

We inspected SAM weights to understand the module's behaviour. SAM weights tended to be lower for slices at the ends of a stack (i.e., first and last slices). For stacks containing at least 3 slices, the mean weight for end slices was about 4.7 times lower (0.28 vs. 1.32) than that for non-end slices. Note that for this comparison, we normalised SAM weights such that a volume of equally important slices each have a weight of 1, to overcome the variation in the number of slices per nodule. Fig. 5 provides examples of this disparity in SAM weights between end and non-end slices. End slices tended to contain smaller cross-sectional areas of the nodule that were less representative of the overall visual characteristics.

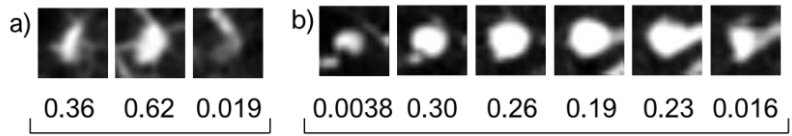

Fig. 5. Sections of transaxial CT images of two nodules. Slice attention weights for each slice are presented. Weights sum to 1 for each nodule; higher values indicate greater importance of that slice.

A performance comparison between the baseline 2D model with and without SAM against the 3D variant is presented in Table 2. The overall error of the 2D baseline without attention was higher than its 3D version, but the performance was superior with SAM.

Table 2. Comparison of performance between the 2D slice-based model against its 3D volume-based version.

| Method | Performance (Mean Absolute Error) | | | | | | | | | |
|---|---|---|---|---|---|---|---|---|---|---|
| | Sub | IS | Cal | Sph | Mar | Lob | Spi | Tex | Mal | Mean |
| 2D Slice-based | 0.741 | 0.031 | 0.228 | 0.542 | 0.610 | 0.483 | 0.450 | 0.498 | 0.510 | 0.455 |
| 2D Slice-based + SAM | 0.658 | **0.026** | 0.165 | **0.489** | **0.549** | **0.460** | **0.423** | **0.409** | **0.457** | **0.404** |
| 3D Volume-based | **0.651** | **0.026** | **0.156** | 0.528 | 0.570 | 0.491 | 0.456 | 0.432 | 0.469 | 0.420 |

We examined the CAAM weights to understand the relative importance of the 9 attributes for each other; these are shown in Fig. 6. The weights of the CAAM + ASCMM variant (Fig. 6c) more closely resembled the relationships exhibited by the correlations between ground truth attribute scores (Fig. 6a) when compared with CAAM only (Fig. 6b). Notable interdependencies include those between shape-related attributes (sphericity, margin, lobulation, and spiculation), and between margin and texture. In contrast, the patterns shown by the CAAM weights of the model without ASCMM largely did not resemble the relationships suggested by the radiologists' scores.

We also assessed the relationships of the CAAM weights between each attribute pair via correlation to discern the similarity between the profiles of the weights. We computed the correlations between the weights of all pairs of attributes per nodule, and visualise the normalised mean correlations in Fig. 6d and e. The relationships shown by the CAAM + ASCMM variant closely mirrored those of the ground truth. While the CAAM-only model identified the associations between margin and texture, and lobulation and spiculation, its resemblance to ground truth was overall weaker. In addition, there were erroneous associations between sphericity and lobulation, and sphericity and spiculation.

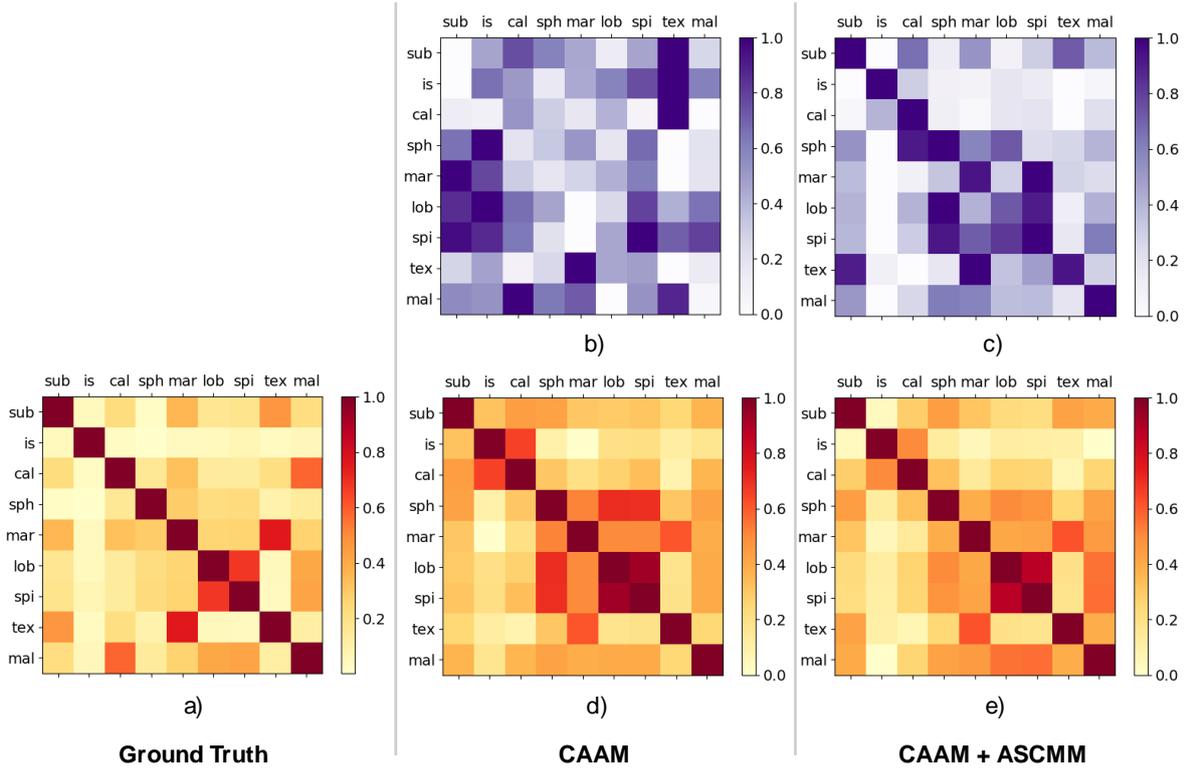

**Ground Truth**    **CAAM**    **CAAM + ASCMM**

Fig. 6. Relationships between attributes. Each entry *ij* in row *i*, column *j* shows the weight (purple) or correlation (orange) of attribute *j* for attribute *i*. a) Correlation magnitudes between radiologists' scores for all attribute pairs. Normalised mean CAAM weights for the baseline model with b) CAAM only and c) both CAAM and ASCMM. Correlations between CAAM weights for all attribute pairs, for the baseline model with d) CAAM only and e) both CAAM and ASCMM. A version displaying the value of each square is included as Fig. S2 in the Supplementary Materials.

The MAEs between the auxiliary and final predictions produced by the variant of the model with CAAM and ASCMM are compared in Table 3. The values indicate a large improvement in performance after further feature processing via the CAAM module.

Table 3. Comparison of performance between auxiliary and final predictions of the baseline model with CAAM and ASCMM.

| Method | Performance (Mean Absolute Error) | | | | | | | | | |
|--------|------|------|------|------|------|------|------|------|------|------|
| | **Sub** | **IS** | **Cal** | **Sph** | **Mar** | **Lob** | **Spi** | **Tex** | **Mal** | **Mean** |
| Auxiliary | 0.681 | 0.062 | 0.354 | 0.495 | 0.567 | 0.475 | 0.437 | 0.423 | 0.462 | 0.440 |
| Final | **0.671** | **0.022** | **0.146** | **0.490** | **0.550** | **0.454** | **0.423** | **0.408** | **0.453** | **0.402** |

## 4.2. Comparison against the State-of-the-Art

Our model, with the attention modules (SAM, ASCMM, and CAAM), had an overall superior performance compared to state-of-the-art methods (Table 4). Note the absence of the MAE for malignancy of MTMR-Net is because the model carried out classification rather than score regression for this attribute. We also modified our model to carry out classification instead of regression for malignancy and achieved competitive performance compared to previous methods (Table 5). We include some example predictions from our model in Fig. 7.

Table 4. Comparison of prediction performance of the proposed model against the state-of-the-art.

| Method | Performance (Mean Absolute Error) | | | | | | | | | |
|---|---|---|---|---|---|---|---|---|---|---|
| | Sub | IS | Cal | Sph | Mar | Lob | Spi | Tex | Mal | Mean |
| IB | 0.84 | **0.02** | 0.20 | 0.85 | 0.85 | 0.79 | 0.67 | 0.46 | 0.88 | 0.62 |
| MTR [11] | 0.65 | 0.06 | 0.33 | 0.67 | 0.72 | 0.76 | 0.70 | 0.53 | 0.70 | 0.57 |
| MTMR-Net [12] | **0.54** | 0.03 | 0.56 | 0.59 | **0.54** | 0.54 | 0.49 | 0.44 | N/A | 0.47 |
| Proposed | 0.66 | **0.02** | **0.14** | **0.48** | **0.54** | **0.46** | **0.42** | **0.40** | **0.45** | **0.40** |

IB = Inter-observer variation computed across all pairs of radiologist ratings per nodule

Table 5. Comparison of benign-malignant classification against the state-of-the-art.

| Method | Performance (%) | | | | |
|---|---|---|---|---|---|
| | Accuracy | Sensitivity | Specificity | Precision | AUC |
| Song et al. 2017 [39] | 84.2 | 84.0 | 84.3 | N/A | N/A |
| Shen et al. 2017 [36] | 87.1 | 77.0 | 93.0 | N/A | 93.0 |
| Xie et al. 2017 [38] | 93.4 | 91.4 | 94.1 | N/A | 97.8 |
| Xie et al. 2019 [34] | 91.6 | 86.5 | 94.0 | 87.8 | 95.7 |
| Xie et al. 2019 [37] | 92.5 | 84.9 | **96.3** | N/A | 95.8 |
| Liu et al. 2020 [12] | 93.5 | 93.0 | 89.4 | N/A | **97.9** |
| Xu et al. 2020 [35] | 92.7 | 85.6 | 95.9 | 90.4 | 94.0 |
| Proposed | **94.7** | **96.2** | 82.9 | **97.8** | 95.9 |

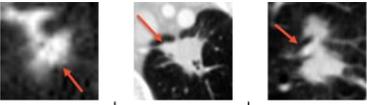

Fig. 7. The middle transaxial CT image slice of three nodules along with predictions from the proposed model. Ground truth (GT) and predicted (P) scores are the left (orange) and right (blue) values in each column, respectively. A smaller error indicates better performance.

# 5. Discussion

Our main findings are that: i) the ablation study showed that our SAM, CAAM, and ASCMM improved the performance of the attention-free baseline model; ii) the attention weights showed that the modules learned to de-emphasise less relevant image slices and exploit meaningful inter-attribute relationships; and iii) our model had superior performance when compared to state-of-the-art methods at scoring the attributes of lung nodules and improved on benign-malignant classification.

## 5.1. Ablation Study

The ablation study revealed the importance of our attention modules in enhancing prediction performance. For SAM, the large disparity between the importance of end and non-end slices in nodule volumes appeared to be related to the representativeness of the slice (Fig. 5). End slices tended to contain fewer nodule pixels, be misleading, and exhibited a stark contrast in subtlety compared to the rest of the stack, often appearing fainter (partly due to the partial volume effect). Hence downweighting end slices for nodules with at least 3 slices improved performance (Table 1). Overall, the results demonstrate that our proposed SAM is an effective module that can proportionally scale the importance of each image slice to filter out non-representative slices and enhance model performance.

We used a 2D CNN as the shared feature extractor backbone of our proposed model. The relative performance of 2D and 3D CNNs for lung nodule analysis in CT shown by previous studies has been mixed, with reports of similar performance, or only slightly better performance from 3D networks [40, 41]. However, 3D CNNs are associated with greater computational expense and increased number of trainable parameters with a higher risk of overfitting. Moreover, 3D CNNs impose a restriction on the depth of the input volume, limiting the compatible range of input nodule sizes. Our model with SAM combines advantages from both approaches in a computationally efficient manner, and it is superior to its 3D counterpart (Table 2). Our model considers information contained in all slices according to their utility as indicated by SAM, without limitations on nodule size and slice thickness, or extra padding to a fixed input dimension.

Analyses of the CAAM weights revealed that the module exploited the inherent relatedness between the 9 attributes, as demonstrated by the similar patterns of inter-attribute relationships compared to ground truth (Fig. 6). With the incorporation of ASCMM into the model, such relationships were better captured and reflective of the inherent inter-attribute relations, translating to greater performance improvement (Table 1). This can be attributed to the auxiliary prediction and loss, which impart a task specialisation effect, as they encourage each attribute branch to be more specialised for its attribute just before CAAM units. The cross-connections are subsequently able to uncover more meaningful interactions and use these interactions for further task refinement between the auxiliary and final predictions (Table 3). This is a distinction of our proposed cross-task attention modules compared to previous cross-connection attention approaches in MTL schemes [21-23]. We also demonstrate that the overall performance is better than training and predicting each attribute individually in Table S2 of Supplementary Materials.

In addition to boosting performance, our attention modules offer analytical and interpretability advantages as they can be used as tools to uncover the underlying complex relationships between the different attributes (Fig. 6). The cause-and-effect connections between attributes for malignancy are particularly useful in the clinical context. Our proposed attention modules are easily interpretable, requiring minimal further computation; the attention weights directly indicate which slices were more influential in prediction outputs and quantifies the strength of relationships between attributes.

The CAAM weights suggest that our model exploited the relatedness between the shape attributes of sphericity, lobulation, margin, and spiculation (Fig. 6c and e). These associations make intuitive sense, e.g. spiculation and lobulation are examples of irregular growth that is a common finding in malignant nodules [42, 43]. Furthermore, the CAAM indicated relationships for malignancy with lobulation, spiculation, margin, sphericity, and texture. These relationships were also indicated by the ground truth correlations and are widely described in literature [4, 5, 16, 17].

## 5.2. Comparison against the State-of-the-Art

Our model is also an MTL framework that leverages the inter-relatedness of all attributes via a shared feature extraction phase, like other state-of-the-art architectures [11, 12]. The previous MTMR-Net model [12] can explicitly explore the relationships between malignancy and the other attributes. Using MTMR-Net, an importance hierarchy of the attributes may be obtained for malignancy. However, this ranking requires an additional logistic regression classifier to be built and extra computation with recursive feature elimination. Our model is unique in its explicit use of all inter-attribute relationships via the proposed CAAM and ASCMM, and is not limited only to those relating to malignancy. Relationships between all attributes are leveraged in a straightforward and intuitive manner using the modules, without needing extra models or complex analyses.

The difficulty of benign-malignant classification is indicated by the incremental improvements of previous models over their antecedent state-of-the-art (Table 5). Our model differs to the compared approaches as it is an MTL framework that also assesses multiple other attributes and does not place a greater priority on malignancy. When repurposed for benign-malignant classification instead of regression, our model performed competitively and improved on overall accuracy (Table 5). Our model also provides the ability to interrogate the pattern of attributes that leads to the overall diagnosis, which is important knowledge in clinical practice.

The strong performance of our attention-free baseline compared to previous slice-based methods (Table 1 and Table 4) suggests that it is effective to assimilate information in all cross-sections of a nodule simultaneously. Individual slices are often not reflective of the entire nodule. Proportionally scaling the influence of each slice on the final predictions according to their usefulness via the proposed SAM further boosts performance, as previously discussed.

The most challenging attributes (aside from malignancy) for our model were subtlety, margin, and sphericity (Table 4). This was consistent with the relative difficulties indicated by the inter-observer variations of the radiologists' scores (IB), and generally by the comparison methods. The high IBs indicate considerable ambiguity in the annotations. For example, for subtlety, nodules that appeared bright and obvious were often rated as 'fairly subtle', while similarly obvious nodules were assigned disparate scores (e.g., contrast Fig. 8a with b). Overall, our model outperformed manual assessment by a large margin.

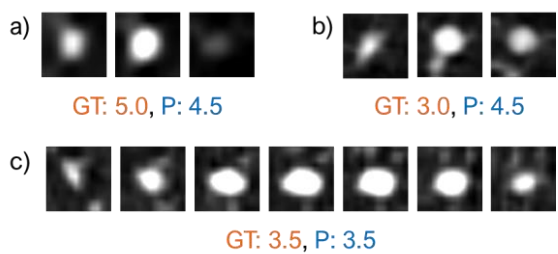

Fig. 8. Sections of transaxial CT images of 3 nodules that exhibit ambiguity in their ground truth score for subtlety. Ground truth (GT) scores are in orange and predicted (P) scores are in blue.

# 6. Conclusion

We proposed a multi-task deep learning model to simultaneously rate the strength of nine attributes of lung nodules on CT images. We introduced three attention modules that improved the performance of our model. Our slice attention module proportionally scales the importance of each cross-sectional slice of a nodule according to its representation of the overall visual characteristics of the nodule. Our cross-attribute feature learning modules explicitly leverage the inter-attribute relationships, exposing such as easily interpretable weights. Our model outperformed state-of-the-art methods at scoring lung nodule attributes.

For future work, the incorporation of nodule context into the model to improve performance on subtlety can be investigated. One possible approach includes characterising visual features of a larger field of view around the target nodule using a separate CNN, then fusing the features with the main model.

In our experiments and those of previous studies [11, 12, 44, 45], nodule ROI localisation was based upon radiologists' annotations, thereby necessitating manual input. Ideally, the localisation pre-processing step should be automated. However, a substantial amount of error will be introduced into the pipeline and be further compounded by inaccuracies from the attribute-scoring model.

We suggest that our model may be extended to other multi-task or multi-modal image-based prediction problems. In particular, our cross-attribute learning attention modules may be leveraged to expose or investigate latent inter-task relationships. Additionally, the interpretable attention weights may be correlated with other data related to the multi-task problem, such as those of genomics.

## Acknowledgements

This work was supported in part by Australian Research Council (ARC) grants (DP170104304 and IC170100022).

# Supplementary Materials for Attention-Enhanced Cross-Task Network for Analysing Multiple Attributes of Lung Nodules in CT

## Fig. S1. High-Resolution Figure of the Proposed Model

To be included as a separate image file.

## Fig. S2. Fig. 6 with Values

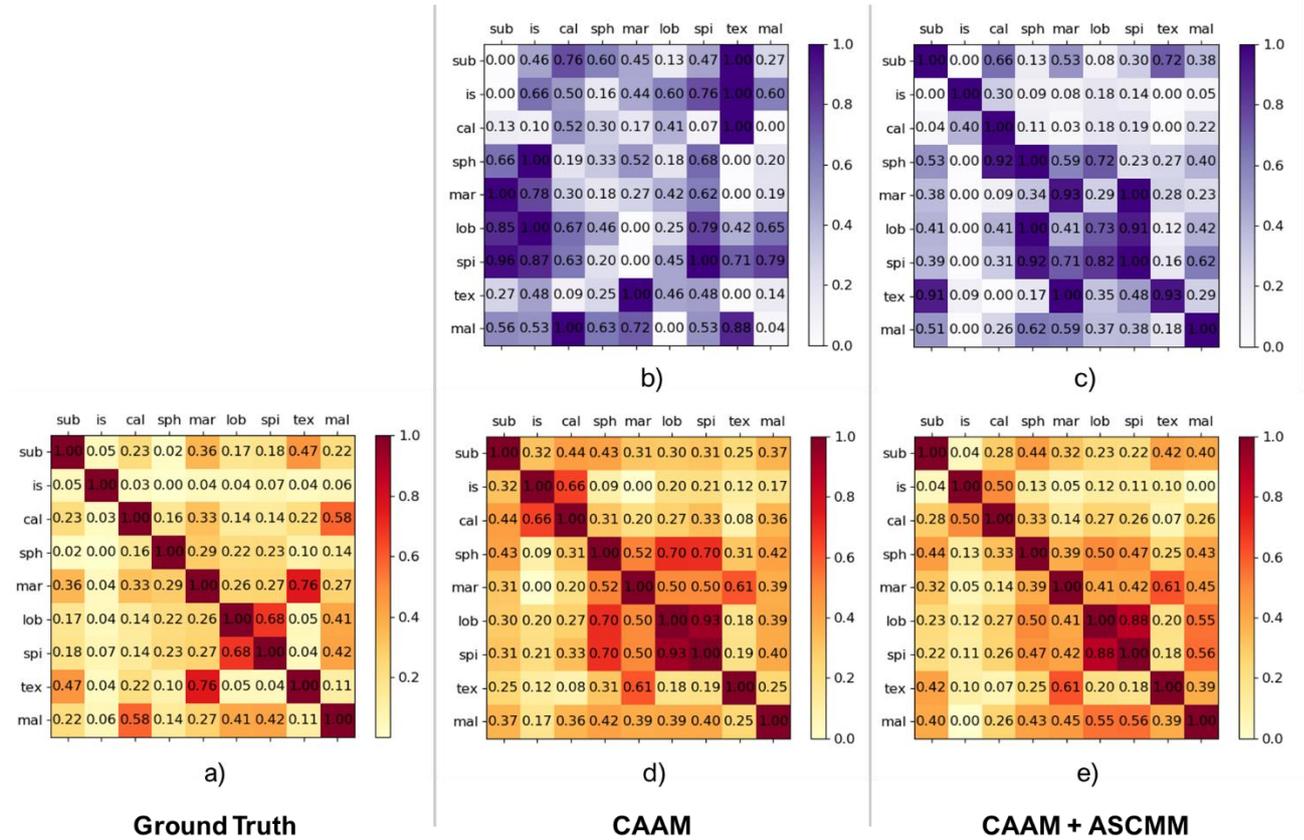

## Table S1. Performance with Varying λ

| λ | Performance (Mean Absolute Error) | | | | | | | | | |
|---|---|---|---|---|---|---|---|---|---|---|
| | **Sub** | **IS** | **Cal** | **Sph** | **Mar** | **Lob** | **Spi** | **Tex** | **Mal** | **Mean** |
| 0.01 | 0.657 | 0.020 | 0.147 | 0.490 | 0.553 | **0.449** | 0.423 | 0.411 | 0.455 | 0.401 |
| 0.1 | 0.664 | 0.020 | 0.141 | **0.484** | **0.536** | 0.460 | 0.424 | **0.400** | 0.453 | **0.398** |
| 0.5 | 0.655 | 0.027 | 0.131 | 0.498 | 0.551 | 0.458 | 0.427 | 0.402 | 0.462 | 0.401 |
| 1 | **0.650** | 0.020 | 0.132 | 0.494 | 0.550 | 0.464 | 0.425 | 0.414 | 0.456 | 0.401 |
| 10 | 0.660 | **0.018** | **0.127** | 0.500 | 0.557 | 0.455 | **0.421** | 0.410 | 0.458 | 0.401 |

**Table S2. Performance of Individually Trained Attributes**

| Method | Performance (Mean Absolute Error) | | | | | | | | | |
|---|---|---|---|---|---|---|---|---|---|---|
| | **Sub** | **IS** | **Cal** | **Sph** | **Mar** | **Lob** | **Spi** | **Tex** | **Mal** | **Mean** |
| Individual | **0.664** | **0.014** | 0.226 | **0.475** | 0.558 | 0.481 | 0.437 | 0.411 | **0.453** | 0.413 |
| Proposed | **0.664** | 0.020 | **0.141** | 0.484 | **0.536** | **0.460** | **0.424** | **0.400** | **0.453** | **0.398** |